\documentclass[11pt]{article}  %Do not change the point size.
\usepackage{menuproc}
\usepackage{cite}
\usepackage{epsfig}
\usepackage{amsmath,amssymb}
\newcommand{\rts}{ \sqrt s}
\def \etal{{\it et al.}}

\begin{document}
%
%____________________________________________________________
%
%  Title, authors, institutions, and abstract
%----------------------------------------------------------------
%  Syntax:  \titlematter{title}{authors}{institutions}{abstract}
%----------------------------------------------------------------
%     If lines are too long, use linebreaks where convenient.
%     If all authors are from the same institution, omit raised letters.
%
\titlematter{Chiral dynamics and pion-nucleon scattering around
 the $N^*(1535)$ resonance}%
{T. Inoue, J.C. Nacher, M.J. Vicente Vacas and E. Oset}%
{Departamento de F\'{\i}sica Te\'orica and IFIC, 
 Centro Mixto Universidad de Valencia-CSIC \\
Institutos de Investigaci\'on de Paterna,
Apdo. correos 22085, 46071, Valencia, Spain}
{ We study here the S-wave interaction of mesons with baryons 
 in the strangeness $S=0$ sector in a coupled channel unitary approach. 
 See ref. \cite{Inoue} for details.
}
%
%
%____________________________________________________________
%  Start article here:
%%%%%%%%%%%%%%%%%%%%%%%%%%%%%%%%%%%%%%%%%%%%%%%%%%%%%%%%%%%%%%%%%%%%%%%%%%%%%%%%%%
\section{Introduction}
 
 The various meson-baryon scatterings are studied 
 in a chiral unitary model with the Bethe-Salpeter equation,
 where the on-shell amplitudes are factorized 
 and the BS equation is reduced to an algebraic equation
 \cite{Kaiser,OllerOsetRamos,Nacher,OllerMeissner,Nieves}.  
 The scattering matrix at the total center of mass energy $\rts$,
 is given by
\begin{equation}
   T(\rts)  = [1 - V(\rts)G(\rts)]^{-1} V(\rts)
   \label{eqn:sol}
\end{equation}
 where $V$ is a transition potential matrix
 and $G$ is a diagonal matrix representing the loop
 integral of a meson and a baryon. 
 The matrix $V$ is taken from 
 the lowest order chiral Lagrangian involving mesons and baryons.
 For example, in the case of S-wave, $V$ is written as
\begin{equation}
 V_{ij}(\rts)
   = - C_{ij}\frac{1}{4f^2}\bar u(p')\gamma^{\mu} u(p) (k_{\mu}+k'_{\mu} )
 \label{eqn:chv}
\end{equation}
 with the meson weak decay constant $f$,
%the initial(final) baryon spinor $u(p)$ ($u(p')$) and 
 and the initial(final) meson momentum $k$ ($k'$).
 The coefficients $C_{ij}$ which reflect the flavor symmetry of the problem,
 are obtained from the Lagrangian.
 The divergent loop integral $G$ defined by 
\begin{equation}
  G_i(\rts) = i
           \int \! \! \frac{d^4 q}{(2 \pi)^4} 
           \frac{2 M_i}{(P-q)^2-M_i^2 + i \epsilon}
           \frac{1}{q^2 - m_i^2 + i \epsilon } 
\label{eqn:oneloop}
\end{equation}
 with $P\equiv(\rts, \vec 0)$ and the baryon(meson) mass $M_i$($m_i$),
 is done with some regularization.
 The finite contribution of the renormalization
 which appears in the real part of $G_i(\rts)$, 
 is treated as an unknown parameter 
 and determined through the fitting to the data.
 The imaginary part of $G_i(\rts)$ is proportional 
 to the phase space and ensures unitarity. 
 For example, with the dimensional regularization, 
 the integral is calculated as
\small
\begin{eqnarray}
 \mbox G_i(\rts) 
 \!\!\!\!\!&=&\!\!\!\!\!  
 \frac{2 M_i}{(4 \pi)^2} 
    \left\{ 
        a_i(\mu) + \log \frac{m_i^2}{\mu^2} + 
        \frac{M_i^2 - m_i^2 + s}{2s} \log \frac{M_i^2}{m_i^2} 
    \right.
  \\
  && \!\!  
      +\frac{Q_i(\rts)}{\rts} 
           \log \left(  s-(M_i^2-m_i^2) + 2 \rts Q_i(\rts) \right) 
      + \frac{Q_i(\rts)}{\rts} 
           \log \left(  s+(M_i^2-m_i^2) + 2 \rts Q_i(\rts) \right) 
  \nonumber
  \\
  && \!\!  
   \left. -\frac{Q_i(\rts)}{\rts}
          \log \left( -s+(M_i^2-m_i^2) + 2 \rts Q_i(\rts) \right) 
          - \frac{Q_i(\rts)}{\rts}
	  \log \left( -s-(M_i^2-m_i^2) + 2 \rts Q_i(\rts) \right)
    \right\}  
  \nonumber
\end{eqnarray} 
\normalsize
 where $a_i(\mu)$ is the contribution of the higher order counter terms
 and $Q_i(\rts)$ is the on-shell center of mass momentum
 of $i$-th meson-baryon system\cite{OllerMeissner}.
 
 To study the pion-nucleon scattering,
 the six coupled meson-baryon systems,
 \{ $\pi^- p$, $\pi^0 n$, $\eta n$, $K^+ \Sigma^-$,
 $K^0 \Sigma^0$, and $K^0 \Lambda$ \}, are considered.
 It is found that qualitatively good S-wave scattering amplitudes are 
 obtained with appropriate values of $a_i(\mu)$ \cite{Nacher}. 
 However, in this simple model, quantitative agreement is not achieved,
 particularly in the isospin 3/2 sector.
 In this paper, we improve this model 
 and try to reproduce the $\pi N$ scattering in a wide energy range.
\section{VMD inspired chiral coefficients and the $\pi \pi N$ channels}
 First, we recall that 
 the S-wave meson-baryon amplitudes from the lowest order chiral Lagrangian 
 are equivalent to the amplitudes of vector meson exchange in the t-channel, 
 in the vector meson dominance(VMD) hypothesis.
 This indicates that the Lagrangian is the 
 effective manifestation of the vector meson exchange mechanism.
 According to this consideration, 
 we introduce the following correction to the chiral coefficient 
 to account for the momentum transfer dependence of the vector meson propagator
\begin{equation}
  C_{ij} \ \rightarrow \
  C_{ij} \times \int
           \! \frac{d \hat k'}{4 \pi} 
           \, \frac{-m_{v}^2}{(k'-k)^2-m_{v}^2} 
  ~~~\mbox{at}~~~ \rts > \sqrt{s_{ij}^{0}}
 \label{eqn:modify}
\end{equation}
 where $\sqrt{s_{ij}^{0}}$ is the energy where the integral of
 (\ref{eqn:modify}) is unity, 
 and which appears in between the thresholds of the two $i,j$ channels. 
 At low energies, this correction is negligible
 but this is not the case at the intermediate energies studied here.
 In fact the $\rho$ meson tail, with $m_{\rho}=770$ MeV, 
 reduces the $\pi^- p \leftrightarrow \pi^- p$ element 
 about 25\% at energies around 1500 MeV.

 Second, we extend our model to include the $\pi \pi N$ channels
 and $\pi N \leftrightarrow \pi \pi N$ transitions.
 We consider the BS equation (\ref{eqn:sol})
 with the eight coupled channels including the six meson-baryon channels
 and two $\pi \pi N$ channels, $\pi^0 \pi^- p$ and $\pi^+ \pi^- n$.
 The $\pi^0 \pi^0 n$ channel is not included because 
 it does not couple to the S-wave $\pi N$.
 We treat the transition potentials of S-wave 
 $\pi N \leftrightarrow \pi \pi N$ as free inputs and determine them 
 so that they account for the data.
 We introduce a two loop integral $\tilde G(\rts)$ 
 for the intermediate $\pi \pi N$ state 
\begin{equation}
  \tilde G(\rts) = i^2 
     \int \! \! \frac{d^4 q_1}{(2 \pi)^4} 
     \int \! \! \frac{d^4 q_2}{(2 \pi)^4} 
      (\vec q_1 - \vec q_2)^2  
      \frac{2 M_N}{(P- q_1- q_2)^2 - M_N^2 + i \epsilon}
      \frac{1}{q_1^2 - m_{\pi}^2 + i \epsilon} 
      \frac{1}{q_2^2 - m_{\pi}^2 + i \epsilon} 
\label{eqn:gtilde}
\end{equation}
 which includes the vertex structure for the factorization. 
 The real part of $\tilde G(\rts)$ in the renormalized model 
 has several subtraction terms to be fixed by the data
 and we find it to be compatible with zero.

\section{Results and discussions}

\begin{figure}[b]
\parbox{.6\textwidth}
  {\epsfig{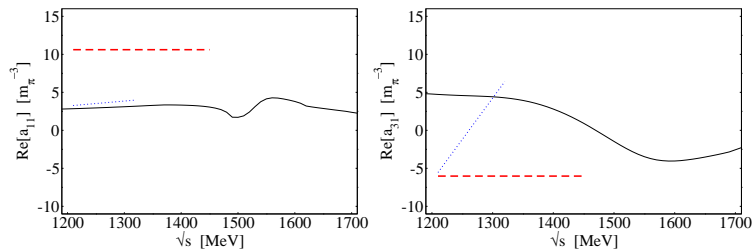}}
\parbox{.37\textwidth}
{\caption{\label{fig:aijre}
 The S-wave $\pi N \leftrightarrow \pi \pi N$ amplitudes
 $a_{11}$ and $a_{31}$. 
 The solid lines are the real part obtained in the present model.
 The dashed and dotted lines are those of 
 \cite{Manley} and \cite{Burkhardt} respectively. }}
\end{figure}

 By varying the input $\pi N \leftrightarrow \pi \pi N$ potential
 and the subtractions in the real part of loop integrals,
 we find a reasonable set of them which 
 reproduce both the elastic $\pi N$ scattering 
 and the pion production cross sections simultaneously.
 The subtraction parameters $a_i(\mu)$ 
 for the meson-baryon loop integrals which we obtain are
\begin{equation}
   \mu= 1200 ~\mbox{MeV},~~ 
   a_{\pi N}(\mu)     =  2.0 ,~~
   a_{\eta N}(\mu)    =  0.1 ,~~
   a_{K \Lambda}(\mu) =  1.5 ,~~ 
   a_{K \Sigma}(\mu)  = -2.8 ~~.
\label{eqn:subpipin}
\end{equation} 
 The determined S-wave $\pi N \leftrightarrow \pi \pi N$
 amplitudes $a_{11}$(isospin 1/2) and $a_{31}$ (isospin 3/2)
 are shown in Fig.\ref{fig:aijre},
 where we compare the real part of them 
 with two empirical ones \cite{Manley,Burkhardt}.
 The lower energy part of the our $a_{11}(\rts)$ agrees with 
 that of the paper \cite{Burkhardt} but it is quite different from
 the one of \cite{Manley}. 
 On the other hand our $a_{31}(\rts)$ amplitude is different from 
 both \cite{Manley} and \cite{Burkhardt}. 
 While it is possible to reproduce the $\pi N \to \pi \pi N$ cross sections
 with the three set of amplitudes, a simultaneous description of the 
 $\pi N \to \pi \pi N$ cross sections and the $\pi N \to \pi N$ scattering date
 is not possible with the amplitude of \cite{Manley} and \cite{Burkhardt}.

\begin{figure}[t]
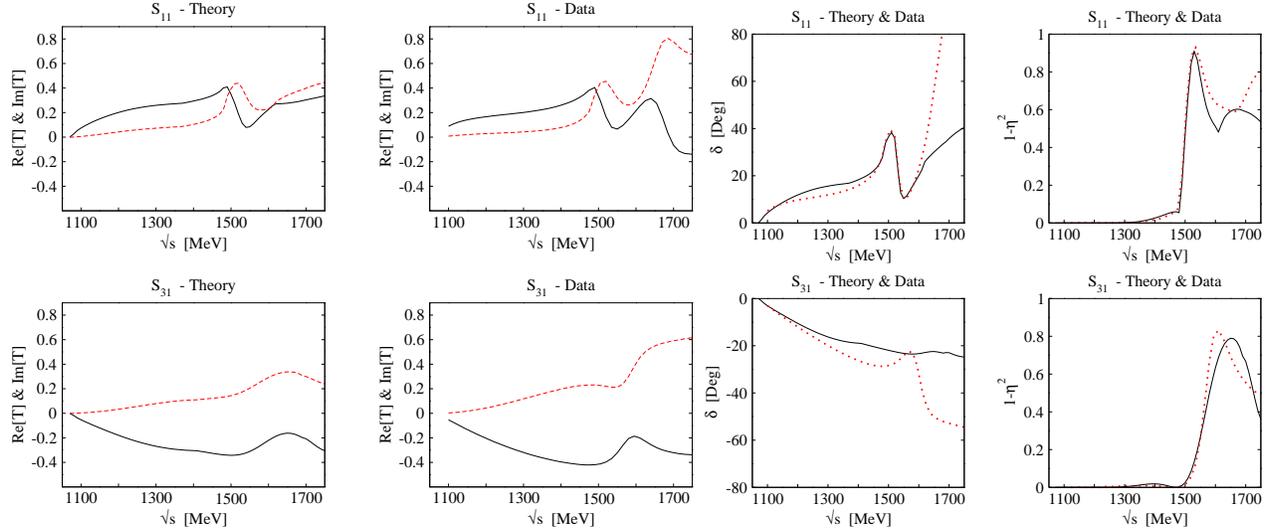

\centerline{
  \epsfig{file= fig15.eps,width=.55\textwidth,silent=,clip=}
  \epsfig{file= fig16.eps,width=.45\textwidth,silent=,clip=}
           }
\caption{\label{fig:tmat}
   Scattering amplitudes, phase-shifts and inelasticities   
   for the $S_{11}$ and $S_{31}$ $\pi N$ partial waves. 
   The solid(dashed) lines in amplitudes are the real(imaginary) parts.
   The dotted lines in the phase-shifts and inelasticities   
   correspond to the data analysis of ref. \cite{CNS}. }
\end{figure}

 The resulting scattering amplitudes, phase-shifts and inelasticities
 for the $S_{11}$ (isospin 1/2) and  $S_{31}$ (isospin 3/2) $\pi N$ partial waves,
 are shown in Fig.\ref{fig:tmat}.
 One can see that results agree with the data in 
 the energy range from threshold to 1600 MeV. 
 We cannot obtain agreement in such a broad range 
 without the correction of the chiral coefficient.
 It shows the importance of the correction.
 Another important thing to note is that, as seen in this figure,
 the inelasticities are well reproduced even at low energies 
 in both $S_{11}$ and $S_{31}$. 
 These are provided by the $\pi \pi N$ channels.
 If $\pi \pi N$ channels are not included,
 the inelasticities of $S_{11} (S_{31})$ below 1487 (1690) MeV 
 are zero and do not agree with data. 
  
%%%%%%%%%%%%%%%%%%%%%%%%%%%%%%%%%%%%%%%%%%%%%%%%%%%%%%%%%%%%%%%%%%%%%%%%%%%%%%%%%%
\acknowledgments{ 
%We would like to thank J.A. Oller and A. Hosaka for useful discussions.
 This work has been partly supported by the Spanish Ministry of Education
 in the program 
 ``Estancias de Doctores y Tecn\'ologos Extranjeros en Espa\~na'',
 by the DGICYT contract number BFM2000-1326
 and by the EU TMR network Eurodaphne, contact no. ERBFMRX-CT98-0169.}
%
%
%%%%%%%%%%%%%%%%%%%%%%%%%%%%%%%%%%%%%%%%%%%%%%%%%%%%%%%%%%%%%%%%%%%%%%%%%%%%%%%%%%
%____________________________________________________________
%  Start references here:


\begin{references}[9]

\bibitem{Inoue}
T. Inoue, M.J. Vicente Vacas and E. Oset, hep-ph/0110333 

\bibitem{Kaiser} 
N. Kaiser, T. Wass, W. Weise, Nucl. Phys. A612(1997)297

\bibitem{OllerOsetRamos} 
J.A. Oller, E. Oset and A. Ramos, Prog. Part. Nucl. Phys. 45(2000)157

\bibitem{Nacher} 
J.C. Nacher, \etal~,
%J.C. Nacher, A. Parreno, E. Oset, A. Ramos, A. Hosaka and M. Oka, 
Nucl. Phys. A678(2000)187

\bibitem{OllerMeissner} 
J.A. Oller and Ulf-G. Meissner, Phys. Lett. B500(2001)263

\bibitem{Nieves} 
J. Nieves and E. Ruiz Arriola, hep-ph/0104307

\bibitem{Manley}
D. Mark Manley, Phys. Rev. D30(1984)536

\bibitem{Burkhardt}
H. Bukhardt and J. Lowe, Phys. Rev. Lett. 67(1991)2622

\bibitem{CNS}
Center of Nuclear Study, http://gwdac.phys.gwu.edu/
\end{references}
\end{document}